\documentclass[12pt]{article}

\newcommand{\be}{\begin{equation}}
\newcommand{\ee}{\end{equation}}
\newcommand{\missingEt}{\mbox{$E_T\hspace{-0.45cm}/$}\ }
\newcommand{\missingpt}{\mbox{$p_T\hspace{-0.45cm}/$}\ }
\begin{document}
\begin{flushright}
YUMS--98/11 \\
September, 1998\\
hep-ph/9903214\\

\end{flushright}
\vskip 5pt
\begin{center}
{\Large \bf Stop Squark Search at Tevatron in the Light Slepton Scenario}\\ 
\vspace*{6mm}
Aseshkrishna Datta $^{a,}$\footnote{E-Mail: asesh@juphys.ernet.in}, 
Monoranjan Guchait$^{b,}$\footnote{E-Mail: guchait@cskim.yonsei.ac.kr}, and 
Kyoung Keun Jeong$^{b,}$\footnote{E-Mail: kkjeong@theory.yonsei.ac.kr} \\
$^a${\it Department of Physics, Jadavpur University, Calcutta - 700032, 
India } \\
$^b${\it Department of Physics, Yonsei University, Seoul 120--749, Korea} \\
\end{center}
\vspace*{1cm}

\begin{abstract}
In minimal supergravity based supersymmetry models the  
charged sleptons 
as well as the sneutrinos may be lighter than the squarks.
In such light slepton scenario the search
for lighter stop squark has been investigated in the dilepton +
missing $p_T(\missingpt)$ +jets ($\ge 1$) channel at Tevatron.
In this scenario semileptonic decay of lighter stop squark has been considered
via off shell and on 
shell charginos. In the latter case the chargino undergoes pure leptonic 2-body
decay. 
We observe that for some favourable region of the MSSM parameter space one can
probe lighter stop squark mass upto 140 GeV with few events  
in the present data set for which the luminosity
is 110 pb$^{-1}$. 
At the MI upgrade one is likely to end up with more events.
\end{abstract}

\clearpage

\noindent
\section*{1. Introduction}
It is well known that supersymmetry (SUSY) \cite{haber} is a very elegant theoretical 
framework which basically invokes a symmetry between fermions and bosons.
The Minimal Supersymmetric extension of the Standard Model(SM), popularly
known as the MSSM, stands out as the most attractive model to go beyond
the SM. Although it cures the SM from many of its ills, one has to pay the
price in terms of a large number of free parameters and a plethora of
predicted new fundamental supersymmetric particles (the SUSY particles or
sparticles). On the face of the fact that the predictions of the SM have been
verified very precisely at recent LEP experiments \cite{lep} and none of these sparticles
has still been discovered it should be kept in mind all along that these
have not contradicted any of the MSSM predictions either. Naturally search
for SUSY has become a thrust area in the phenomenological studies and 
experiments at the present and future generation of colliders.

Based on the gauge group $SU(3)_c \otimes SU(2)_L \otimes U(1)_Y$ alike
the SM, the MSSM contains ordinary SM 
particles and their  
corresponding superpartners which differ by spin 1/2.  
In order to generate masses for both up and down type quarks by 
electroweak symmetry breaking 
the MSSM requires at least two $SU(2)$ higgs doublets.
After electroweak symmetry breaking one is left with  
five physical higgses --- two charged ($H^{\pm}$) higgses, 
two neutral CP even $(h, H)$ higgses and one  
CP odd (pseudoscalar) higgs($A$) \cite{higgs}.
The supersymmetric partners of the charged and the neutral SM gauge bosons,
called gauginos, are also not the physical states. They get mixed up because 
of electroweak symmetry breaking leading to   
physical supersymmetric states (the mass eigenstates)   
called charginos($\tilde W_i; i=1,2)$ 
and neutralinos($\tilde Z_j;j=1,2,3,4$). 
The charginos and the neutralinos are the linear compositions of 
charged($\tilde W^{\pm},\tilde H^{\pm}$) and neutral 
($\tilde W_3, \tilde B, \tilde {H^{0}}_1, \tilde {H^0}_2$) 
gauginos respectively\cite{haber}. 
In this paper we shall assume $\tilde W_1$ as the lighter chargino and 
$\tilde Z_1$ as the lightest neutralino. In most of the models the lightest
neutralino, $\tilde Z_1$ is assumed to be the lightest supersymmetric
particle(LSP). In models with conserved R-parity \cite{weinberg}  
the LSP turns out to be stable and all other heavy sparticles eventually
cascade to the LSP. Since the LSP must be  neutral and weakly interacting
on cosmological ground 
it leads to missing 
energy which is the characteristic SUSY signal. 
Also, the left and right handed SM fermions have their respective spin zero 
supersymmetric states called 
sfermions viz. $\tilde f_L$ and $\tilde f_R$. 
The mixing between the chiral sfermions are 
proportional to the masses of their corresponding SM partners. 
Therefore, such mixing becomes important only for the third generation 
of sfermions. 

In the last few years, extensive searches for SUSY were carried out 
at LEP\cite{susylep} as well as Tevatron\cite{cdfsusy} 
experiments. Non-observation of any 
SUSY signal excluded certain region of the MSSM parameter space 
and put lower bounds on different sparticle masses. 
However, it is to be noted that these searches were 
performed  under some specific assumptions in the models and  
on the masses of the sparticles. 
In most of the cases these assumptions made the analyses very simple 
but were in no way  
compelling. Moreover,these restrict a variety of other 
phenomenological possibilities in the context of SUSY searches.
In this work we have 
addressed one such issue in the context of stop squark search at 
Tevatron.

Due to the generic large top Yukawa coupling, the third generation of squarks
viz. the stop squark, undergoes substantial mixing between the left($\tilde t_{L})$ and 
the right (${\tilde t_R}$) chiral states \cite{rudaz}. 
As a consequence, one of the physical 
states($\tilde t_1$) may be lighter than all  
other squarks in the MSSM.    
If this be the actual case and also that the mass of 
$\tilde t_1$($m_{\tilde t_1}$) is less than the gluino  
(the SUSY partner of the gluon) mass ($m_{\tilde{g}}$)  
then $\tilde{t}_1$ can decay into various channels: 
$\tilde t_1 \rightarrow (i) c\tilde Z_1$, (ii) $b\tilde W_1$, 
$(iii) b \ell \tilde\nu$, $(iv) b \tilde\ell \nu$, $(v) b W\tilde Z_1$, 
$(vi) b f \bar f' \tilde Z_1$, $(vii) {\tilde b} W$, $(viii) t {\tilde Z_1}$ 
within the MSSM. 
The dominant decay channel of $\tilde t_1$ is mostly determined by  
$m_{\tilde t_1}$ and by the masses of the final state particles. 
In the case of $m_{\tilde t_1} < m_t, m_{\tilde b}$, the decay modes  
(vii) and (viii) are kinematically forbidden. 
The decay mode (vi) consists of four body final states where as the decay
mode (v) involves massive particles in the final state and it is also 
propagator suppressed with top quark in the propagator.   
Therefore, whenever any one (or more) of the 
channels (i) -- (iv), which are primarily two or three body decays,  
opens up that(those) will dominate over other multi-body decay processes 
viz,(v)--(viii)\cite{hikasa}.  
On the other hand, if the masses of the sleptons($m_{\tilde\nu}, m_{\tilde \ell}$)
are much larger than those of $\tilde W_1$ and  $\tilde Z_1$,  
the decay modes (iii) and (iv) are suppressed heavily 
in comparison to the decay modes (i) and (ii). 
Therefore, within this scenario where all other sparticles, except LSP,are much larger
heavier than $\tilde t1_1$,
the latte has only two accessible decay modes viz. either (i) or (ii).
If $m_{\tilde t_1}> m_{\tilde W_1} + m_b$ then the charged current decay 
\be
\tilde t_1 \to b \tilde W_1
\label{eq:tw1}
\ee
will be the dominant one. Otherwise it will decay via loops \cite{hikasa}
with flavour changing neutral currents (FCNC) \cite{hikasa} 
\be
\tilde t_1 \to c \tilde Z_1
\label{eq:cz1}
\ee
and the cross section of such a process is naturally very small. 
In the latter case the stop pair production will
lead to jets plus missing energy carried by the LSP signal\cite{tata}.  
In the former case 
( Eq.~\ref{eq:tw1}), 
the chargino will decay via $W$ (real or virtual)
as 
\be
\tilde W_1 \to \tilde Z_1 f \bar f'
\label{eq:w2z}
\ee 
under the assumption that the sleptons and the squarks  
are heavy. Therefore, if $\tilde W_1$ 
decays leptonically, its decay branching ratio(BR) is essentially 
the BR of $W$ in the leptonic channel which is 2/9 for both $e$ and $\mu$.
This leptonic decay of $\tilde W_1$ leads to 
missing energy plus single lepton or   
unlike sign dileptons\cite{tata}.  
The decay 
channel of $\tilde W_1$ as given by Eq.~\ref{eq:w2z} 
is assumed to be the only decay process of $\tilde W_1$ in 
most SUSY searches. 
Hereafter we shall call those as conventional SUSY searches.   
In theoretical studies where stop squark searches were 
carried out either directly\cite{tata,rohini} or indirectly\cite{mahlon} in the 
context of Tevatron, the decay patterns of $\tilde t_1$ and $\tilde W_1$ are 
assumed to be identical  as we have described above.

If the mass hierarchy among sparticles is such that the 
sneutrinos($\tilde\nu$)~\cite{drees} are much heavier than the 
LSP ($\tilde Z_1$) only and 
lighter than all other sparticles then $\tilde W_1$ 
dominantly undergoes two body decay as
\be
\tilde W_1 \to \ell {\tilde\nu_{\ell}}
\label{eq:w1snu}
\ee
in contrast to Eq.3, 
with  a BR 2/3 where $\ell = e$ and  $\mu$. 
Note that this two body BR of $\tilde W_1$ 
is  larger than that for the leptonic decay mode of Eq.~\ref{eq:w2z}  
by a factor of 3. 
If the associated left handed charged slepton ($\tilde\ell_L$)
is also lighter than $\tilde{W}_1$ then $\tilde W_1$ can also
decay in the two body channel  
\be
\tilde W_1 \to \tilde\ell_L \nu_{\ell}
\label{eq:w1sl}
\ee
followed by 
\be
{\tilde\ell_L} \stackrel{100\%}{\longrightarrow} \ell {\tilde Z_1}
\label{eq:cdk}
\ee 
In this light sneutrino scenario, the sneutrino 
decays in the invisible channel  
$\tilde\nu \to \nu \tilde Z_1$ with 100 \% BR whenever $\tilde Z_1$ 
has some amount of gaugino content and thus leads to missing energy 
in the final state. Therefore, in addition to 
$\tilde Z_1$, $\tilde\nu$ will behave like a LSP in such a scenario. 
Interestingly, it is seen that when $m_{{\tilde\nu}({\tilde l_L})} +m_b 
<m_{\tilde t_1}<m_{\tilde W_1} + m_b$, $\tilde t_1$ 
will decay via off-shell $\tilde W_1$ either in the channel 
(iii) or (iv)~\cite{hikasa} i.e. 
\be
\label{eq:t1blsnu}
\tilde t_1 \to b \tilde W_1^* \to b(\ell \tilde\nu_{\ell})
\ee
or
\be
\label{eq:t1bslnu}
\tilde t_1 \to b \tilde W_1^* \to b(\tilde\ell_L \nu_{\ell}).
\ee
These three body decay modes of $\tilde t_1$
via off-shell $\tilde W_1$ always dominate over the loop level 
decay mode (i). Hence, most of the time the 
pair production of stop squark leads to \emph{unlike sign dilepton plus jets accompanied 
by missing energy} due to the presence of $\tilde\nu$ 
and $\tilde Z_1$. In the above we have put subscript L in $\tilde \ell$ to 
imply that gauge dominated $\tilde W_1$ interacts with left handed 
sleptons only.

In Ref.\cite{tata}, a study on stop squark search has been  
carried out in the $1\ell$ $+$ $b+jets$ + $\missingEt$  and     
$\ell^{+}{\ell^{'}}^{-}$ $+$ jets + $\missingEt \ $  channels for the 
leptonic decays
of $\tilde W_1$(Eq.~\ref{eq:w2z}) when $m_{\tilde t_1} > m_b + m_{\tilde W_1}$ 
In this scenario appreciable jet activitys
are expected as $\widetilde{W}_1$ decays hadronically (about 80\%).  
On the other hand, in the light 
slepton scenario (where we can have sleptons lighter than $\tilde{t}_1$)     
stop squark will always decay semileptonically with less jet activities 
as $\tilde W_1$ will not have any hadronic decay mode (Eqs. 4---8). 
As a consequence, the events containing only single lepton will be 
absent and the      
jets + $\missingEt$  
signal will be degraded.  
Indeed, in Ref.\cite{tata} the authors have mentioned the possibility 
of enhancement of the dilepton signal for light sleptons. 
As we have mentioned 
above, this enhancement (by a factor of 10) of signal 
events occurs due to the eventually larger BR for the leptonic decay of $\tilde W_1$.  
On the other hand, in our scenario leptons are comparatively
softer when $\tilde t_1$ undergoes the two body decay (Eq.~\ref{eq:tw1}) whereas
leptons are harder in the case of its three body decay via off shell
$\tilde W_1$. 
In view of these facts it is instructive to study the prospect of 
stop searches systematically for the relevent MSSM parameter space in 
this light slepton scenario.  
To the best of our knowledge, so far there is no such study at Tevatron
energies, although experimental searches for $\tilde t_1$ has already been 
carried out at LEP in this scenario~\cite{stop}. 
In our analysis we have taken into account the presently available 
experimental bounds on the relevant MSSM parameters. 
The phenomenological impacts of the light slepton scenario in the context of
SUSY searches have been discussed in detail in a series of works\cite{guchait}. 
This work adds to that series.

The paper is organised as follows. In section 2 we describe 
very briefly the stop squark masses and mixing  and also present 
relevant formulae for the decay width of stop squark. 
In section 3 we  discuss our results.

\section*{2. Relevant formulae}

As already mentioned in the introduction, due to the large Yukawa interactions,
the left ($\tilde t_L$) and the right ($\tilde t_R$) handed stop squarks  
get mixed up\cite{rudaz}. 
The stop mixing matrix 
in the basis ($\tilde t_L, \tilde t_R$) is expressed as 
\be
\label{eq:stm}
M^{2}_{\tilde t} = \pmatrix{m^{2}_{\tilde t_L} & a_t m_t \cr
a_t m_t & m^{2}_{\tilde t_R}} 
\ee
where
\begin{eqnarray}
m^{2}_{\tilde t_L} & = & m_{\tilde Q_3}^2 + m_t^2 + 
(\frac{1}{2} - \frac{2}{3} \sin^{2}\theta_W)M_Z^2 \cos 2\beta \nonumber \\
m^{2}_{\tilde t_R} & = & m_{\tilde U_3}^2 + m_t^2 + 
\frac{2}{3} M_Z^2 \sin^{2}\theta_W \cos 2\beta \nonumber \\
a_t & = & (A_t + \mu \cot\beta ) \nonumber 
\end{eqnarray}
and $\tan \beta$, $\mu$ and $A_t$ are the ratio of vacuum expectation 
values($\tan\beta$ = $v_1/v_2$) of two higgs doublets, 
the SUSY higgs mass parameter and the trilinear
coupling respectively. The soft mass terms for the third generation of 
doublet, $m_{\tilde Q_3}$ and the up type singlet, $m_{\tilde U_3}$ are 
related to that corresponding to the first/second generation of squarks as

\begin{eqnarray}
\label{eq:sqm}
m_{\tilde Q_3}^2 = m_{\tilde Q_1}^2 - I \nonumber \\
m_{\tilde U_3}^2 = m_{\tilde U_1}^2 - I
\end{eqnarray}
where $I$ is a function of Yukawa coupling determined by 
the renormalization group equation in minimal supergravity models  
assuming GUT scenario.
The physical stop squark states $\tilde{t_1}$ and $\tilde{t_2}$ are related to  
the chiral states as  
\be
\pmatrix{\tilde t_1 \cr \tilde t_2} = 
\pmatrix{\cos\theta_{\tilde t} & - \sin\theta_{\tilde t} \cr
\sin\theta_{\tilde t} & \cos\theta_{\tilde t} } 
\pmatrix{\tilde t_L \cr \tilde t_R}
\ee
where the mixing angle $\theta_{\tilde t}$ is given by 
\be
\label{eq:thet}
\tan 2\theta_{\tilde t} = \frac{2 a_t \, m_t}{m_{\tilde t_R}^2 - 
m_{\tilde t_L}^2 }
\ee
and the mass eigenvalues are given by 
\be
\label{eq:mt12}
m_{\tilde t_{1,2}}^2 
= \frac{1}{2} \left[ m_{\tilde t_L}^2 + m_{\tilde t_R}^2
\mp \left\{ (m_{\tilde t_L}^2 - m_{\tilde t_R}^2)^2 
+ {(2a_t m_t)}^2 \right\}^{1/2} \right] .
\ee
The diagonal terms in the stop mass matrix of Eq.~\ref{eq:stm} are very small 
because of Eq.~\ref{eq:sqm} where they receive the large negative
contribution from $I$ which depends on the large top Yukawa coupling.  
If we notice the expression for the mass eigenvalues (Eq.~\ref{eq:mt12}),
it is quite obvious that if the SUSY parameters are of the same 
order of magnitude then $m_{\tilde t_1}$ would be very light and it may even 
be lighter than the top quark unlike other squark masses.
Eq.~\ref{eq:thet} implies that the mixing angle is 
proportional to the mass of the top quark which in turn implies that 
such mixings for other generation of squarks are very small due to 
their small masses.
In the following we discuss very briefly about the 
possible decay modes of the lighter stop   
squark($\tilde t_1$) ((i) -- (iv)) which are relevant in our scenario.  

\bigskip

{\bf (i)} $\mathbf{\tilde{t_1} \to c \tilde{Z}_1 \;\; :}$
It is the only allowed decay mode of $\tilde{t}_1$ when $m_{\tilde t_1}$ is 
lighter than all other sparticles except the LSP.  
It is shown\cite{hikasa} that 
this flavour changing decay proceeds through various loops 
which are logarithmically divergent. 
The logarithmic part of these loop diagrams induce the mixing 
between $\tilde c_L$, $\tilde t_L$ and $\tilde t_R$ whereas  
$\tilde c_R$ does not mix with $\tilde t_L$ and $\tilde t_R$ 
in the limit $m_c \rightarrow 0$. 
The corresponding decay width of $\tilde{t_1}$ in this flavour changing  
mode is 
\be
\Gamma(\tilde {t_1} \to c \tilde Z_1) = \frac{g^2}{16\pi} |\epsilon|^2 f^2
m_{\tilde t_1} \left( 1-\frac{m_{\tilde Z_1}^2}{m_{\tilde t_1}^2} \right)^2
\label{eq:tcg}
\ee
where
\[ f = \sqrt{2} \frac{N_{12}}{\cos\theta_W} \left( \frac{1}{2} 
- \frac{2}{3} \sin^2\theta_W \right) + \frac{2\sqrt{2}}{3} N_{11} 
\sin \theta_W \]
$N_{11}, N_{12}$ being the $\tilde B$ and $\tilde W_3$  
components of $\tilde Z_1$ state\cite{uv}.
Here $\epsilon$ is the $\tilde c_L$ - $\tilde t_L$ mixing parameter
and is significantly very small ($\simeq 10^{-4}$)\cite{hikasa}. This leads to
a moderate decay width $\Gamma (\tilde{t}_1 \to c \tilde{Z}_1)$.
Therefore, the other decay modes (ii)--(iv),  
whenever kinematically allowed, will dominate over the present mode.  

\bigskip
{\bf (ii)} $\mathbf{\tilde{t_1} \to b \tilde{W}_1 \;\;:}$ 
Here the two body decay of $\tilde t_1$ proceeds through 
tree level charged current interaction when  
$m_{\tilde t_1}> m_b + m_{\tilde W_1}$. The decay width is
\begin{eqnarray} 
\label{eq:st2bwino}
\Gamma (\tilde t_1 \to b \tilde W_1) 
&=& \frac{\alpha}{4\sin^2\theta_W} m_{\tilde t_1}\; \lambda^{1/2}
 \left( 1, \frac{m_b^2}{m_{\tilde t_1}^2} ,
 \frac{m_{\tilde W_1}^2}{m_{\tilde t_1}^2} \right) \nonumber \\
&\times& \left [\{ |c_L|^2 + |c_R|^2 \}
\left( 1 - \frac{m_b^2}{m_{\tilde t_1}^2}
 - \frac{m_{\tilde W_1}^2}{m_{\tilde t_1}^2} \right)
 -\frac{4 m_b m_{\tilde W_1}}{m_{\tilde t_1}^2} \mbox{Re}(c_L c_R^*) \right] \\
c_L &\equiv& - \frac{m_b \; U_{12}}{\sqrt{2} m_W \cos\beta}
 \cos\theta_{\tilde t} \nonumber \\
c_R &\equiv& V_{11} \cos\theta_{\tilde t}
 + \frac{m_t V_{12} \sin\theta_{\tilde t}}{\sqrt{2} m_W \sin\beta} \nonumber
\end{eqnarray}
where $U$, $V$ are the chargino mixing matrix\cite{uv} corresponding to 
the right and the left handed states respectively. 
Since it is a charged current two body decay process, 
it will decay dominantly with 100\% BR whenever it is kinematically 
allowed. 

\bigskip
{\bf (iii)} $\mathbf{\tilde {t_1} \to b \ell \tilde \nu_{\ell} \;\; :}$ 
This decay takes place via off shell $\tilde W_1$(decay mediated by off 
shell $\tilde W_2$ is suppressed due to its large mass) 
and dominates when $m_{\tilde{\nu}} + m_b < m_{\tilde t_1} < m_b + m_{\tilde\ell_L} < 
m_b + m_{\tilde W_1}$ and the decay width is 
\be
\label{eq:st2snu}
\Gamma(\tilde t_1 \to b \, \ell \tilde\nu_{\ell}) =
\frac{(\alpha V_{11})^2}{16 \pi \sin^4\theta_W m_{\tilde t_1}} 
\int W(x_b,x_\ell) dx_\ell dx_b
\ee
where
\begin{eqnarray}
W(x_b,x_\ell) &=& \frac{1}{(1 + \mu_b - x_b - \mu_{\tilde W_1})^2} \nonumber \\
&\times& \bigg[ c_L^2 m_{\tilde W_1}^2 
        (1 + \mu_{\tilde\nu} - \mu_b - 2 + x_b + x_\ell) \nonumber \\
&+& 2 c_L c_R m_{\tilde W_1} m_b (1 + \mu_b - \mu_{\tilde\nu}- x_b) 
        \nonumber \\
&+& c_R^2 m_{\tilde t_1}^2 (1 - \mu_b - \mu_{\tilde\nu}- x_\ell)
        (1 + \mu_b - \mu_{\tilde\nu}- x_b) \nonumber \\
&-& c_R^2 m_{\tilde\nu}^2 (1 + \mu_{\tilde\nu} - \mu_b - 2 + x_b + x_\ell) 
\bigg]
\end{eqnarray}
where $\mu$'s and $x$'s are defined below. Here we have neglected the 
masses of the SM leptons.   

\bigskip
{\bf (iv)} $\mathbf{\tilde{t_1} \to b \nu_{\ell} \tilde{\ell_{L}} \;\; :}$ 
This decay occurs through off shell $\tilde W_1$ and dominates when
$m_{\tilde\ell_L} + m_b < m_{\tilde t_1} < m_b + m_{\tilde W_1}$ 
and the decay rate is given by 
\be
\label{eq:st2sl}
\Gamma(\tilde t_1 \to b \,\tilde \ell_{L} \nu_{\ell}) =
\frac{(\alpha U_{11})^2}{16 \pi \sin^4\theta_W m_{\tilde t_1}} 
\int W(x_b,x_{\tilde \ell})dx_{\tilde \ell} dx_b
\ee
where
\begin{eqnarray}
W(x_b,x_\ell) &=& \frac{1}{(1 + \mu_b - x_b - \mu_{\tilde W_1})^2} \nonumber \\
&\times& \bigg[ c_L^2 {m_{\tilde t_1}^2} (1 + \mu_b - \mu_{\tilde \ell}- x_b)
(1 - \mu_b - \mu_{\tilde \ell}- 2 + x_b + x_{\tilde \ell} ) \nonumber \\
&-& c_L^2 m_{\tilde \ell}^2 (1 + \mu_{\tilde \ell} - \mu_b - x_{\tilde \ell})
+ 2 c_L c_R m_{\tilde W_1} m_b (1 + \mu_b - \mu_{\tilde \ell}- x_b) 
        \nonumber \\
&+& c_R^2 m_{\tilde W_1}^2 (1 + \mu_{\tilde \ell} - \mu_b - x_{\tilde \ell}) 
\bigg] 
\end{eqnarray}

The couplings $c_L$ and $c_R$ are the same as in Eq.~\ref{eq:st2bwino} and
\begin{eqnarray}
\mu_b = m_b^2/m_{\tilde t_1}^2 & \mu_{\ell(\tilde\ell)} = 
m_\ell^2/m_{\tilde t_1}^2
        & \mu_{\tilde\nu} = m_{\tilde\nu}^2/m_{\tilde t_1}^2 \\
x_b = 2E_b/m_{\tilde t_1} & x_{\ell(\tilde\ell)} = 2E_\ell/m_{\tilde t_1}
        & x_{\tilde\nu} = 2E_{\tilde\nu}/m_{\tilde t_1} \\
\mbox{with} & x_b + x_{\ell(\tilde\ell)} + x_{\tilde\nu(\nu)} = 2 &
\end{eqnarray}
The range of integration of the above Eq.~\ref{eq:st2snu}, \ref{eq:st2sl}
are\cite{bp} 
\begin{eqnarray}
&{}& 2\sqrt{\mu_b} \le x_b \le
 1 + \mu_b - \mu_c - \mu_d - 
2\sqrt{\mu_c \mu_d}  \label{domain1} \\
&{}& \frac{ (2 - x_b)(1 + \mu_b + \mu_c - \mu_d - x_b)
- \sqrt{(x_b^2 - 4\mu_b) \lambda(1 + \mu_b - x_b, \mu_c, \mu_d)} }
{2(1 - x_b + \mu_b)}
\le x_c  \nonumber \\
&{}& \le \frac{ (2 - x_b)(1 + \mu_b + \mu_c - \mu_d - x_b)
+ \sqrt{(x_b^2 - 4\mu_b) \lambda(1 + \mu_b - x_b, \mu_c, \mu_d)} }
{2(1 - x_b + \mu_b)} \label{domain2}
\end{eqnarray}
where $c = \ell (\tilde \ell)$ and $d = \tilde \nu(\nu)$ for Eq.~\ref{eq:st2snu}
(~\ref{eq:st2sl}).

It is not difficult to understand the behaviour of the above two three-body decay modes 
of $\tilde t_1$ (Eq.~\ref{eq:st2snu} and Eq.~\ref{eq:st2sl} (if both are 
kinematically allowed)) qualitatively by simply examining the dimension of 
the interaction Lagrangians for both the processes\cite{hikasa}. 
It is to be noted that chargino does not interact with the right handed 
charged sleptons if one neglects the masses of corresponding SM leptons.
The interaction Lagrangians for these two decay modes 
contain terms like $\tilde t_L \bar b_L l_L \tilde\nu_L^*$ for (iii) 
and $\tilde t_L \bar b_L \bar\nu_l \tilde l_L$ for (iv). 
In the former case one needs a $\gamma$ matrix to contract 
the chirality between the fermion and the antifermion and that has to 
be, in turn,  contracted with the derivative of the scalar field. 
Therefore, on dimensional ground, a coefficient $\sim m^{-2}$ 
is necessary in this term. On the other hand, in the charged slepton case 
the operator does not require any derivative and it has dimension five, 
and hence it requires only a coefficient $\sim m^{-1}$\cite{hikasa}.  
Hence the decay mode $\tilde{t_1} \to b \ell \tilde{\nu}$ (Eq.~\ref{eq:st2snu}) 
is always suppressed by an extra mass dimension in comparison to 
the charged slepton decay mode viz. $\tilde{t_1} \to b \nu \tilde{\ell}$ 
(Eq.~\ref{eq:st2sl}). 
Naturally, whenever the latter mode (Eq.~\ref{eq:st2sl})  
is open it dominates over the former (Eq~\ref{eq:st2snu})  
in this scenario.
We have checked this by calculating numerically the decay widths as given by 
Eqs.~\ref{eq:st2snu} and \ref{eq:st2sl}.

In the MSSM, the masses of charginos and neutralinos 
and their corresponding mixing matrix elements $U$, $V$ and $N$  
can be determined by $M_2$, $\mu$ and
$\tan\beta$. Here $M_2$ is the $SU(2)$ gaugino mass 
and it is related to $M_1$, the $U(1)$ gaugino mass, in the following manner  
at the electroweak scale\cite{m1m2} under
the assumption of gaugino mass unification at the GUT scale ----
\be
M_1 = \frac{5}{3} M_2\tan^2{\theta}_W 
\ee
where $M_1$ is the $U(1)$ gaugino mass.
There is also experimental lower bound on the mass of the lighter chargino from the
non observation of SUSY events at LEP experiments. 
Interestingly, limit on the chargino mass is also related
to $m_{\tilde\nu}$. In the case of chargino pair 
production through $e^{+} e^{-}$ collision, in addition to 
the $s$-channel diagrams, there is also a $t$-channel diagram (via 
$\tilde\nu$ exchange) which interferes destructively with the $s$-channel 
and hence, reduces the cross section. 
When $m_{\tilde\nu}$ is larger ($\ge$ 200 GeV) 
than $m_{\tilde W_1}$ the effect is less pronounced. 
But, for $m_{\tilde\nu} < m_{\tilde W_1}$ 
the chargino pair production cross section will reduce 
and also the leptonic branching ratio of $\tilde W_1$ 
(Eq.~\ref{eq:w1snu}) will get enhanced and finally will reduce
the limit of $m_{\tilde W_1}$. The present available 
lower limit of lighter chargino obtained by LEP experiments 
at $\sqrt{s}=$ 160 GeV and 172 GeV\cite{alephc,delphic} are   
\begin{eqnarray}
\label{eq:w1limit}
m_{\tilde W_1} > 85~\mbox{GeV} &\mbox{for}& m_{\tilde\nu} > 200~\mbox{GeV}
 \nonumber\\ 
  >67~\mbox{GeV} &\mbox{for}& 41~\mbox{GeV} < m_{\tilde\nu}<100~\mbox{GeV}
\end{eqnarray}
It is to be noted that the bound on lighter chargino in the case of 
light sneutrino holds when mass difference between them is greater 
than 10 GeV. Therefore, for almost degenerate sneutrino and lighter 
chargino with the lighter chargino heavier than the sneutrino there is no  
such bound. The only relevent bound on chargino mass then comes 
from LEP-1 which is little better than $M_Z/2$.  
As in our present analysis we consider the case of light sneutrino 
($m_{\tilde\nu} < 100~\mbox{GeV}$),
the only relevant limit on $\tilde W_1$ is the second one.
In the stop sector, we have used $m_{\tilde t_1}$ 
and $\theta_{\tilde t}$ as input parameters.  
The ALEPH collaboration\cite{stop} at LEP has come about with a lower
bound on $m_{\tilde t_1}$
\be
m_{\tilde t_1} >  70~\mbox{GeV}
\label{eq:stb}
\ee
analysing the data taken upto $\sqrt{s} = 172$~GeV in  
the channel ${\tilde t_1}\rightarrow b\ell \tilde\nu$ 
assuming a mass difference between the $\tilde t_1$ and $\tilde\nu$ 
of at least 10 GeV. This bound is independent of 
the mixing angle $\theta_{\tilde t}$ in the stop sector.
On the other hand, D\O\ collaboration at Fermilab have also constrained 
$m_{\tilde t_1}$--$m_{\tilde Z_1}$ plane analysing their data for 
$\tilde t_1\rightarrow c \tilde Z_1$ in  
$jets + \missingEt$ channel \cite{d0stop}.
As described above, those D\O\  constraints on $m_{\tilde t_1}$ 
will not work in our analysis. They have also carried out search for 
stop squark 
\cite{d0stopee} in the dielectron channel assuming the conventional 
cascade decay of stop(see, Eq.~\ref{eq:tw1,w2z}). But effectivley, no limit on 
$m_{\tilde t_1}$ is set.  
The other very crucial parameter of our analysis is $m_{\tilde\nu_{\ell}}$. 
The light sneutrino scenario can be easily accommodated if 
one embeds GUT in the MSSM. Further, with the assumption of a common scalar
mass ($m_0$) at the GUT scale 
the slepton masses get related\cite{slept} at the electroweak scale  
and the slepton masses at the electroweak scale as ---- 
\begin{eqnarray}
m_{\tilde \ell_R}^2 &=& m_0^2 + 0.22 M_2^2 - \sin^2\theta_W M_Z^2
        \cos 2\beta \nonumber \\
m_{\tilde \ell_L}^2 &=& m_0^2 + 0.75 M_2^2 - 0.5(1-2\sin^2\theta_W) M_Z^2
        \cos 2\beta \\
m_{\tilde\nu_{\ell}}^2 &=& m_0^2 + 0.75 M_2^2 + 0.5 M_Z^2 \cos 2\beta \nonumber
\label{eq:slep}
\end{eqnarray}
where $\ell$ is $e$ or $\mu$. In case of staus, the mixing plays an 
important role for very large $\tan\beta$ and $\mu$ parameter. 
In some region of the $m_0$, $M_2$ and $\tan\beta$ parameter space 
the masses of charged sleptons($\tilde \ell_L, \tilde \ell_R$) and 
also that of $\tilde\nu$ may be lighter than all the gauginos except the LSP. 
It happens due to the mass-splittings in the presence of $SU(2)$ breaking 
$D$-terms. It is to be noted that the above relations
hold in a model independent way as long as $SU(2)_L$ is a good symmetry 
at the electroweak scale. The light slepton masses are much more natural  
if one allows \emph{non-universal} soft breaking masses\cite{parida} 
and in that case 
our light slepton scenario holds in a larger region of the allowed parameter space.

The ALEPH collaboration\cite{aleph} has carried out direct searches for  
right handed selectron($\tilde e_R$) and smuon 
($\tilde \mu_R$) 
in the opposite sign dilepton plus missing energy channel 
at energies $\sqrt s =$ 161 GeV and 172 GeV. 
From the non-observation of events they have obtained lower limits on 
$m_{\tilde e_R}$ and $m_{\tilde\mu_R}$.
The absolute values of the limits depend on the relative mass 
differences between the sleptons and the LSP. 
Since the slepton masses are determined by two parameters 
$M_2$ and $m_0$ for a given $\tan\beta$, the limits on 
slepton masses can be translated so as to constrain the $m_0$--$M_2$ plane.  
In Ref. \cite{aleph} the excluded region in the  
$m_0$--$M_2$ plane has been shown for $\mu = -200$~GeV and $\tan\beta = 2$.  
In our analysis we have used this constrained $m_0$--$M_2$ parameter 
space to determine $m_{\tilde\nu_{\ell}}$ and $m_{\tilde\ell_L}$
which are subsequently used as inputs to our analysis.

\section*{3. Results and Discussion}

At Tevatron the dominant mechanisms for stop production are the leading
order QCD processes like quark-antiquark annihilation and gluon-gluon fusion\cite{sally}
\be
q \bar q \to \tilde t_1 \bar{\tilde t_1} ~;~ gg \to \tilde t_1 \bar{\tilde t_1}
\ee 
The next to leading order(NLO) corrections to these processes 
has been computed recently\cite{spira}. 
This correction is only a few percent and is positive. 
In our conservative estimate we have not taken into account this 
NLO correction.
The cross section for stop pair production depends only 
on $m_{\tilde t_1}$ and its dependence on other MSSM parameters
is almost negligible even it does not depend on $\theta_{\tilde t}$ 
as it is mainly a QCD process. 
So, any bound obtained on the stop production cross section 
from experiment can be translated to a  bound on 
the stop mass in a straight forward manner. 
In our calculation 
we have used the CTEQ3L\cite{pdflib} for parton density setting the 
QCD scale at $m_{\tilde t_1}$.     

Once a pair of stop squarks is produced at the Tevatron,  
depending on its mass, the stop squark will decay through either of the 
channels (ii)--(iv) as described in section 1, essentially with 
100\% branching ratio in the light slepton scenario.
If $m_{\tilde t_1} < m_{\tilde W_1} + m_b$, 
then $\tilde t_1$ will decay 
either via the channel in Eq.~\ref{eq:t1blsnu} 
or in Eq.~\ref{eq:t1bslnu} depending 
on $m_{\tilde \nu}$ and $m_{\tilde \ell}$,
as discussed in the last section.
If $m_{\tilde t_1} > m_b + m_{\tilde W_1}$,
then $\tilde t_1$ will decay through channel $\tilde{t_1} \to b 
\widetilde{W_1}$ 
(Eq.~\ref{eq:tw1}) and subsequently $\tilde W_1$ will decay as indicated in 
Eq.~\ref{eq:w1snu} or Eq.~\ref{eq:w1sl}.
As a consequence, pair production of ${\tilde t_1}$ eventually leads to 
a signal consisting of \emph{opposite sign dilepton(${\ell^{+}}{\ell^{-}}$) 
and jets(mainly $b$-jets) along with missing transverse momentum  
($\missingpt $)} due to the presence of 
$\tilde\nu$ and/or $\tilde Z_1$. 
The dilepton event from $t \bar{t}$
pair production is topologically similar to this and hence acts as the  
dominant SM background to this signal. 
The other sources of the  
SM backgrounds are $W$-pair production and the Drell-Yan processes. 
We have mentioned in the introduction that the signal, consisting 
same final states
for stop squarks at the Tevatron, although in a different scenario, has been 
studied in Ref.\cite{tata} where cascade decays of $\tilde W_1$ 
are considered as in Eq.~\ref{eq:w2z}. 
In that analysis the authors optimised  a set of kinematic 
cuts by which the SM backgrounds can be minimised without 
much affecting the signal rates. 
In our parton level analysis we have used those  
optimised cuts\cite{tata}, viz. 
\begin{enumerate}
\item $p_T^{\ell}> 10~\mbox{GeV}$, $|\eta_\ell| < 1 $ 
 and $E_{AC}^T < $10\% $p_T^\ell$
\item $20^o < \phi_{\ell^+ \ell^-} < 160^o$
\item $\not{p_T} > 25~\mbox{GeV}$
\item $B < 100~\mbox{GeV}$ 
 where $B = p_T^{\ell_1} + p_T^{\ell_2} + \not{p_T}$
\item number of jets, $n_j \ge 1$ with $p_T^j > 15~\mbox{GeV}$ 
 and $|\eta_j| < 2$
\end{enumerate}
The lepton selection cut (1) has been applied to both the leptons. 
The cut (2) removes the unlike sign dileptons due to the Drell-Yan 
process where lepton pairs emerge back to back most of the time.
We have already mentioned that the most significant background comes from 
$t \bar t$ and $WW$ productions where the leptons come from real $W$'s.  
For the signal process, the leptons come either from the 3-body decay of
$\tilde t_1$ (see Eqs.7-8) or from the 
2 body decay of $\tilde W_1$(see Eqs.4-6). 
Hence the leptons  are always accompanied by a massive
particle which is either a $\tilde\nu$ or a $\tilde Z_1$.  
Therefore, it is expected that the  
leptons in the signal are softer than that from the backgrounds. 
Therefore, by imposing cut on a newly constructed kinematic variable like 
``$B$'', which is the scalar sum of the transverse momenta of two leptons 
and $\missingpt $, one can cope with the backgrounds efficiently. 
Obviously, in the case of signal, leptons are distributed towards
the lower values of ``$B$'' whereas in the case of background 
the situation is just the opposite.
Hence, putting an upper cut(4) on ``$B$'' can effectively reduce the background 
without killing the signal significantly\cite{tata}.
Cut (5), which is the requirement of at least one jet with $p_T$
greater than 15 GeV, effectively suppresses the Drell-Yan background. 
The estimated cross sections for the SM backgrounds from $t\bar t$ and $WW$ 
production under all these cuts as described above 
are 14~fb for $m_t=$170 GeV and 10~fb respectively\cite{tata} and  
we have used these numbers in our analysis.  
Note that the value of $m_t$ conforms with recent experiments\cite{ttop}.

As for an illustration, we have computed the signal cross section 
at $\sqrt s = 1.8$~TeV
for a set of input parameters given in Table 1.
In our analysis we have fixed $\theta_{\tilde t}= - 45^o$, $\mu = -200$~GeV.
The signal rate is not too sensitive to $\theta_{\tilde t}$
since it neither affects the cross section nor the BR 
of $\tilde t_1$ decay which is always nearly 100\% for a particular decay 
channel. This is so because out of various allowed decay channels for 
$\tilde{t}_1$ only one dominates overwhelmingly over the others at a time. 
We have estimated the signal cross section for different $m_{\tilde{t}_1}$ 
ranging from 80--170 GeV keeping in mind its lower bound
from LEP(Eq.~\ref{eq:stb}). 
If $m_{\tilde t_1} > m_t$ then the decay mode (viii), as described in 
section 1, will open up and consequently the signal will be different.
The hardness of the lepton momentum depends on the relative mass differences  
$\Delta m_{\tilde\nu} = m_{\tilde W_1} - m_{\tilde\nu}$ (for $\tilde t_1$
decay via on shell $\tilde W_1$) or 
$\Delta m_{\tilde\ell} = m_{\tilde\ell} - m_{\tilde Z_1}$. 
The larger is the value of this mass difference, harder are the leptons. 
In our analysis we considered $ee$, $\mu \mu$ and $e \mu$ dilepton 
final states i.e. 40\% of stop pair production events will have    
final states with dileptons.

In Fig.1 we have shown the $p_T$ distribution of  
the lepton for some representative values of  
($m_{\tilde t_1}, m_{\tilde\nu}, m_{\tilde W_1}$) = A(105,69,90)~GeV; 
B(105,81.4, 90)~GeV; C(130,69,90)~GeV; D(130,81.4,90) GeV.
The distributions obtained are subject to the kinematic cuts 
as described above except for the $p_T$ cuts on the lepton pairs. 
For a fixed $m_{\tilde t_1}$, the $p_T$ of lepton 
in case of point A is harder than in case of point B
since $\Delta m_{\tilde\nu}$  
is larger for the former. Similar behaviour exists for the points C and D.
As far as the distributions corresponding to A and C are concerned, although 
$\Delta m_{\tilde\nu}$'s are the same,  
the distribution C takes over the distribution A in the higher $p_T$ 
region.  
This enhancement is due to large $m_{\tilde t_1}$ which has boosted 
the lepton in case C. But this behaviour does not 
show up in cases B and D where the extra boost due to large
$m_{\tilde t_1}$ for the point D does not help as the lepton are very much  
less energetic.

It is obvious from Fig. 1 that cut (1) on lepton will
have significant bearings over case B and D and is less significant for 
A and C.   
  
In Fig.2 the signal cross sections for different $m_{\tilde t_1}$ values
are presented for the first four sets of parameters of Table 1. 
For each of the curves the signal cross section 
goes up with increasing $m_{\tilde t_1}$
as long as $\tilde t_1$ is light  
despite a decrease in stop pair production cross section.   
This is because the cut 
efficiency factor goes up for larger $m_{\tilde t_1}$ as the 
leptons and jets become harder for a given set of input parameters. 
What actually happens in such a region is that 
the reduction of 
cross section is overcompensated for by the enhancement in the efficiency of 
the kinematic cuts.  
As for example, for the set (a) 
the stop pair production cross section  
decreases by a factor of 2 as     
$m_{\tilde t_1}$ goes from 80 to 90 GeV, whereas the cut 
efficiency factor increases by an even larger factor. 
It happens since for larger $m_{\tilde t_1}$ harder are the 
$b$-jets and the leptons. 
The `dip's occur in cases (a) and (d) just where 
on shell $\tilde W_1$'s are produced from stop decay 
for some values of $m_{\tilde t_1}$ (see Eq.~\ref{eq:tw1}), while small
$m_{\tilde t_1} - m_{\tilde W_1}$ leads to  
very soft $b$-jets and the signal gets severely affected by the jet selection 
cut. However, such `dip' also appears in the curves for 
the sets (b) and (c) but with negligibly  small cross sections and we have 
not shown it. Note that the the signal cross section dominates over the 
background (24~fb) 
for stop masses upto about 135 GeV for a less favourable region of the pa
space 
and nearly 150 GeV for the most favourable one 
having few events for the present integrated
luminosity of 110~$\mbox{pb}^{-1}$. Also, it is not possible to probe 
stop mass in the region of parameter space where the `dip' occurs. 
It is to be noted that in the conventional scenario 
stop mass up to 100~GeV can be probed with the same luminosity 
option\cite{tata} with few events.
In our scheme, the signal will be viable 
when $m_{\tilde t_1} - m_{\tilde \nu}$ (for off shell $\tilde W_1$) or 
$m_{\tilde t_1}- m_{\tilde W_1}$ (for on shell ${\tilde W_1}$) 
is larger than at least 10 GeV.  
Notice that
for small $m_{\tilde t_1}$ ($\le$ 100 GeV) with $m_{\tilde t_1} >> 
m_{\tilde{\nu}}$,
if this scenario is kinematically allowed, 
it will give a few 100 events(see the case for (a) in Fig.2) for the present
luminosity .
We want to emphasise that if this light slepton scenario is at all instrumental 
for the cascade decays of stop squark then it may turn out as one of its 
viable discovery channels
at Tevatron although over a rather limited  
region of the MSSM parameter space. 
The signal size has also a modest dependence on the MSSM parameters through
$m_{\tilde\nu}$, $m_{\tilde W_1}$ and $m_{\tilde Z_1}$.
But $m_{\tilde W_1}$ does not vary so much with $\mu$ in the gauge 
dominated region i.e. when $M_2 \gg \mid\mu|$. 
Hence, signal size is not that sensitive to $\mu$ parameter. 
However, in the negative $\mu$ region 
$m_{\tilde W_1}$ decreases with the increase of $\tan\beta$.

In Fig.3 we have shown the signal cross section with the conventions similar 
to Fig.2 but for $\tan\beta = 10$. 
We have computed the signal rate for the set of
points (e) and (f) of Table 1. 
Although for higher $\tan \beta$ the constrained region in the $m_0$--$M_2$ 
plane is not available from slepton search at LEP \cite{aleph}, 
still one can qualitatively argue that the value which we have used 
is very much likely to be allowed by the present lower limits on 
the mass of the right handed sleptons. This is  
because the increase of $\tan\beta$ for a given $m_0$ and $M_2$ lowers 
the value of $m_{\tilde\nu}$ whereas it enhances $m_{{\tilde e}_L}$
and $m_{{\tilde\mu}_R}$. The signal rate for set (e) 
decreases in comparison to set (a) since in the former case     
$\Delta m_{\tilde\nu}$ is smaller. On the other hand, for higher 
$\tan\beta$, the signal in case (f) is stronger than that in set (c). 
In case (c), leptons may also come from the charged slepton decays  
(Eqs. 5-6), which involve further cascading compared to case (f) and hence  
may become soft.  
For this higher $\tan \beta$ case the signal
rate is above the background level for $m_{\tilde t_1}$ upto
about 130~GeV.  

In the Main Injector (MI) upgrade the integrated luminosity is expected to 
be enhanced by a factor of 20 which will result in more number of events.
But, unfortunately, the discovery range of $m_{\tilde t_1}$ is not expected
to extend as the signal in this channel is highly limitted by the backgrounds.  
The message from this work is that one can constrain the relative 
differences among $m_{\tilde{t_1}}$, $m_{\tilde{\nu}}$ and $m_{\tilde{W_1}}$
for light $\tilde{t_1}$ simply by analyzing the present dilepton data from
Tevatron in the light slepton scenario. 
Especially, for lower values of $m_{\tilde t_1}$,  
the difference $m_{\tilde t_1} - m_{\tilde\nu}$ can be constrained
analysing the same set of data whence  
one can put a lower bound on $m_{\tilde\nu}$ if the limit of $m_{\tilde t_1}$ 
can be found from any other experiments, say, from the LEP or vice versa.

In conclusion, we have investigated the decay patterns of the light stop
when the sleptons are lighter than it and the charginos as well.
We have studied the prospect of stop
search at Tevatron in such a scenario although the latter is viable only for an 
additionally constrained, but still significant, region 
of the MSSM parameter space. We observe that light stop
masses upto 140 GeV can be probed in some favourable region of this  
parameter space with few events. 
Moreover, it is possible to constrain the light slepton scenario 
in the minimal supergravity (mSUGRA) inspired MSSM from the non-observation 
of any signal in this channel in the present data set at Tevatron.

\section*{Acknowledgements} 

The authors are indebted to Amitava Datta for
many valuable suggestions and careful reading of the manuscript. AD acknowledges
financial assistance from the Department of Science and Technology, Government of India.
MG was
supported in part by a 1997 foreign posdoctoral fellowship through the
Korean Science and Engineering Foundation, and in part by a graduate 
school fellowship from Yonsei University, Seoul.

\newpage
\section*{Figure Captions}
\smallskip
\begin{enumerate}
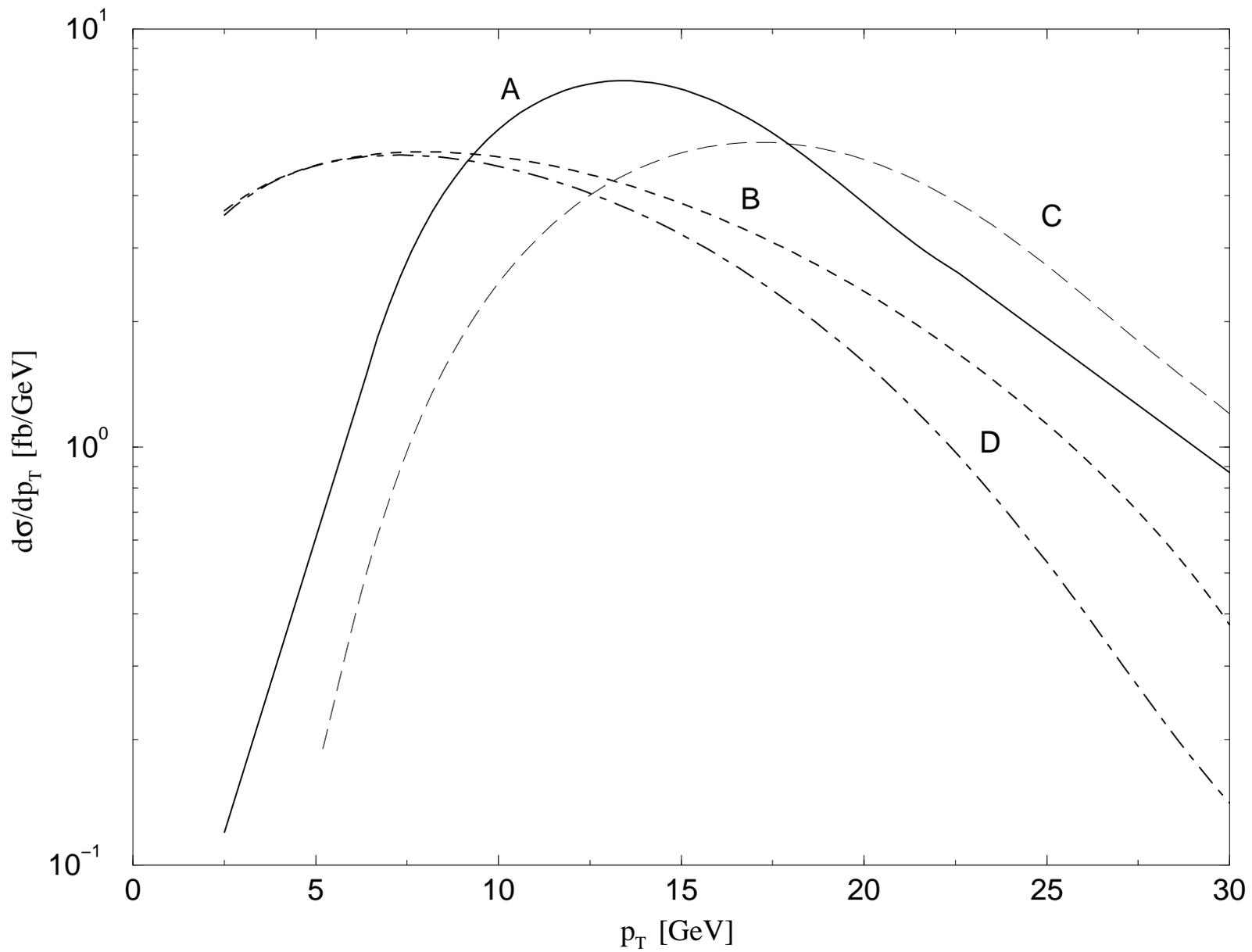
\item[{Fig.1}]: The $p_T$ distributions for one of the leptons are 
shown for the values of 
($m_{\tilde t_1}, m_{\tilde\nu}, m_{\tilde W_1}$) = A(105,69,90)~GeV;
B(105,81.4, 90)~GeV; C(130,69,90)~GeV; D(130,81.4,90)~GeV.  
In case of A and C ,($m_0, M_2$)=(55, 75)~GeV  
whereas for B and D is ($m_0,M_2$)=(70,75)~GeV.  
For all the cases $\tan \beta$ = 2, $\mu =-$200~GeV and 
$\theta_{\tilde t}$ =$ -45^\circ$.

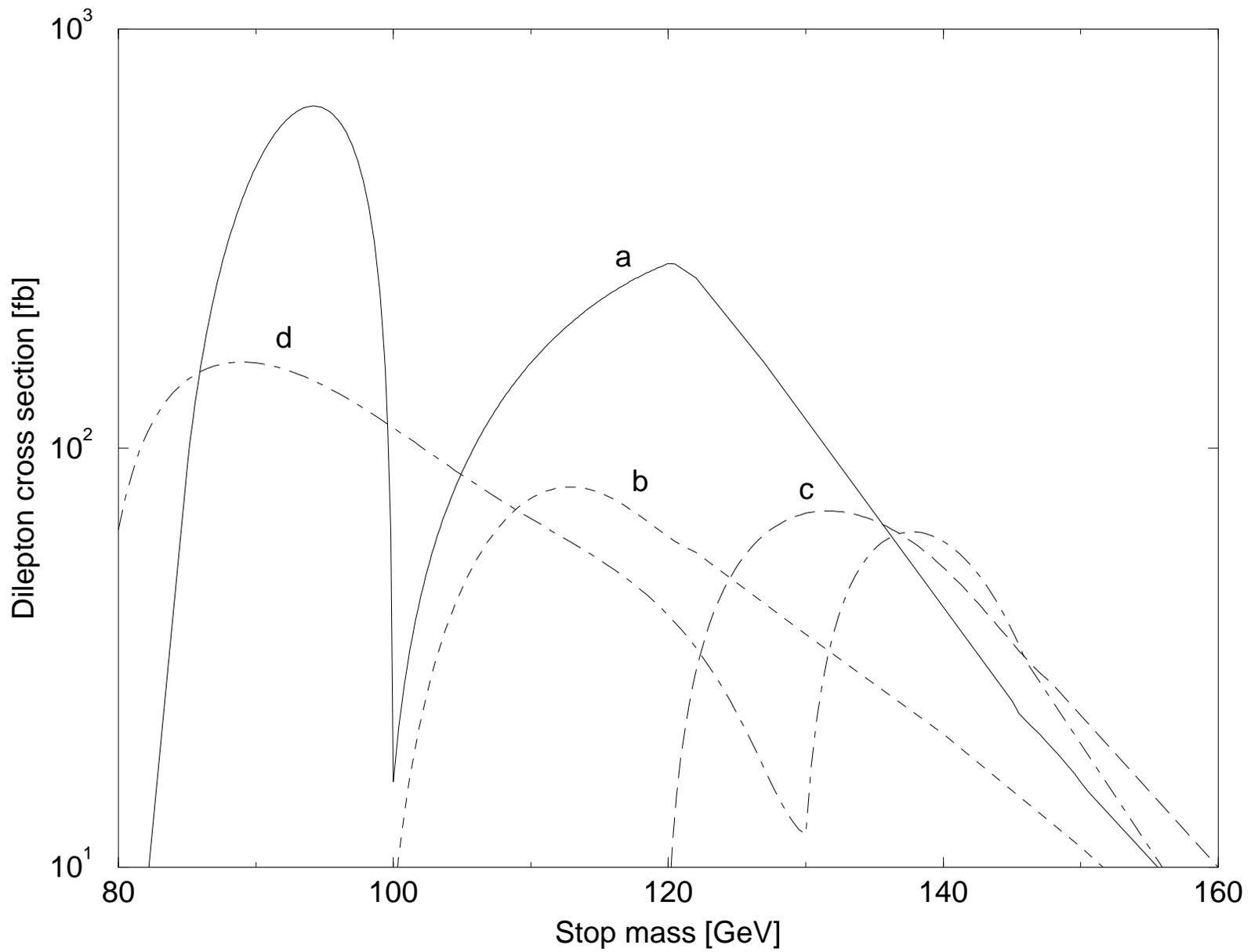
\item[{Fig.2}]: The variations of dilepton signal cross sections are 
shown for different stop
masses at the Tevatron for $\sqrt{s}$=1.8 TeV. The different labels 
a,b,c,d correspond to different sets of parameters as described in Table 1. 
The values of $\tan \beta$ and $\theta_{\tilde t}$ are the same as in Fig.1.

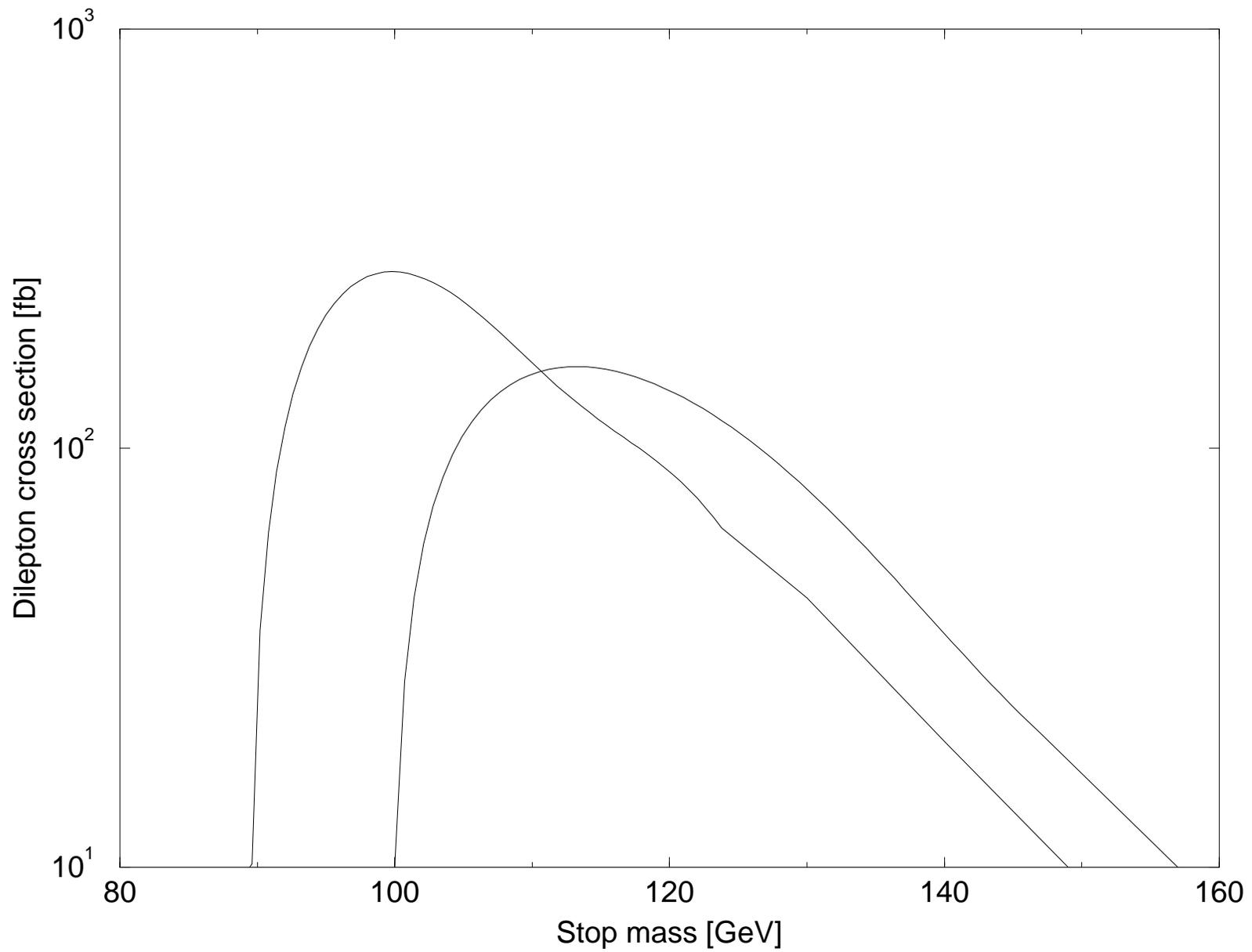
\item[{Fig.3}]: Same as in Fig.2 except for the parameter sets which are (e) 
and (f) of Table 1. 
\end{enumerate}

\vskip 20pt
\begin{table}[hbt]
\begin{center}
{\bf Table 1} \\
\vspace{0.2in}
\begin{tabular}{|c|c|c|c|c|c|c|c|}
\hline 
Set & $m_0$ & $m_2$ & $\tan\beta$ & $m_{\tilde\nu}$ & $m_{\tilde\ell_L}$
 & $m_{\tilde W_1}$ & $m_{\tilde Z_1}$ \\ 
&\hspace{1.5cm} & \hspace{1.5cm} & \hspace{1.5cm} & \hspace{1.5cm}
 & \hspace{1.5cm} & \hspace{1.5cm} & \hspace{1.5cm} \\[-1.2em]
\hline \hline
(a) & 55 & 75 & 2 & 69 & 92.6 & 90 & 41.79 \\
\hline
(b) & 70 & 75 & 2 & 81.4 & 102.28 & 90 & 41.79 \\
\hline
(c) & 40 & 100 & 2 & 86.6 & 102.19 & 112.38 & 54.56 \\
\hline
(d) & 67 & 50 & 2 & 60 & 86.28 & 68 & 28.56 \\
\hline
(e) & 55 & 75 & 10 & 56.35 & 97.16 & 70 & 38 \\
\hline
(f) & 40 & 100 & 10 & 71 & 106.28 & 90.92 & 50 \\
\hline
\end{tabular}
\vspace{0.2in}
\caption{Masses( in GeV) of sneutrino($\tilde \nu$), left handed charged
slepton($m_{\tilde {\ell}}$), lighter chargino ($m_{\tilde W_1}$) and
LSP ($m_{\tilde Z_1}$) are shown for different values of $m_0$, $M_2$ and
$\tan\beta$. The masses of sleptons have been computed using 
Eq.~\ref{eq:slep}. The extreme left column labels different sets of parameters
alphabatically.}  
\end{center}
\end{table}


\newpage
\begin{thebibliography}{99}

\bibitem{haber} For reviews see e.g. H. E. Haber and G. L. Kane, \emph{Phys. Rep.}
{\bf 117}, 75(1985); H. P. Nilles, \emph{Phys. Rep.} {\bf 111}, 1 (1984).

\bibitem{lep} See, for example, A. Altarelli, R. Barbieri, and 
F. Caravaglios, \emph{Int. J. Mod. Phys.} {\bf A13}, 1031, 1998  
(and references there in).

\bibitem{higgs} J. F. Gunion and H. E. Haber, \emph{Nucl. Phys.} {\bf B272}, 1(1986).

\bibitem{weinberg} S. Weinberg, \emph{Phys. Rev.} {\bf D26}, 287 (1982); 
N. Sakai
and T. Yanagida, \emph{Nucl. Phys.} {\bf B197}, 83 (1982); S. Dimopoulos, S. Raby
and F. Wilczek, \emph{Phys. Lett.} {\bf B212}, 133 (1982).

\bibitem{susylep} C. Caso et. al, \emph{Eur. Phys. J.} {\bf C3} 
1 (1998).

\bibitem{cdfsusy} CDF Collaboration, F. Abe et al., \emph{Phys. Rev.} {\bf D56}, 
1357 (1997); 
D\O\ Collaboration, S. Abachi. et al., \emph{Phys. Rev. Lett.} {\bf 75},
618 (1995).

\bibitem{rudaz} J. Ellis and S. Rudaz, \emph{Phys. Lett.} {\bf 128B}, 248 (1983); 
G. Altarelli and R. Ruckl, \emph{Phys. Lett.} {\bf 144B}, 126 (1984).

\bibitem{hikasa} K. Hikasa and M. Kobayashi, \emph{Phys. Rev.} {\bf D36}, 724 (1987).

\bibitem{tata} H. Baer, J. Sender and X. Tata, \emph{Phys. Rev.} {\bf D50}, 4517 (1994).

\bibitem{rohini} H. Baer, M. Drees, J. Gunion, R. Godbole and X. Tata,
\emph{Phys. Rev.} {\bf D44}, 725 (1991). 

\bibitem{mahlon} G. Mahlon and G. L. Kane, \emph{Phys. Rev.} {\bf D55}, 2779 (1997);
S. Mrenna and C. P. Yuan, \emph{Phys. Lett.} {\bf B367}, 188 (1996).

\bibitem{drees} See, e.g. L. E. Ibanez and G. G. Ross, \emph{Phys. Lett.} {\bf B110},
215 (1982); L. E. Ibanez and J. Lopez, \emph{Nucl. Phys.} {\bf B233}, 511 (1984);
M. Drees and M. M. Nojiri, \emph{Nucl. Phys.} {\bf B369}, 54 (1992); N. Polonosky and 
A. Pomarol, \emph{Phys. Rev.} {\bf D51}, 6532 (1995);
Y. Kawamura, H. Murayama and M. Yamaguchi, \emph{Phys. Rev.} {\bf D51}, 1337 (1995)

\bibitem{stop} ALEPH Collaboration, R. Barate et al., \emph{Phys. Lett.} {\bf B413}, 
431 (1997).  

\bibitem{guchait} A. Datta, B. Mukhopadhyaya and M. Guchait, 
\emph{Mod. Phys. Lett.}, {\bf 10}, 1011 (1995); A. Datta, M. Drees and M. Guchait, 
\emph{Z. Phys.} {\bf C69}, 347 (1996); S. Chakrabarty, A. Datta and M. Guchait, 
\emph{Z. Phys.} {\bf C68}, 325 (1995); A. Datta, Aseshkrishna Datta, S. Raychaudhuri, 
\emph{Phys. Lett.} {\bf B349}, 113 (1995); 
A. Datta, Aseshkrishna Datta, S. Raychaudhuri, \emph{Eur. Phys. J.},
{\bf C1}, 375 (1998).

\bibitem{uv} See, Haber and Kane in Ref.[1].

\bibitem{bp} V. D. Barger and R. J. N. Phillips, {\it Collider Physics}, 
Addison-Wesley, 1987.

\bibitem{m1m2} For a micro review see \emph{Supersymmetry} by H. Haber, 
in Review
of Particle Properties, M. Barnett et al. \emph{Phys. Rev.} {\bf D54}, 1 (1996). 

\bibitem{alephc} ALEPH Collaboration, R. Barate et al., CERN PPE/97-128. 

\bibitem{delphic} DELPHI Collaboration, P. Abreu et al., \emph{Eur. Phys. J.}  
{\bf C1}, 1, 1998.

\bibitem{d0stop} D\O\ Collaboration, S. Abachi et al., \emph{Phys. Rev. 
Lett.} {\bf 76}, 
2222 (1996).

\bibitem{d0stopee} D\O\ Collaboration, S. Abachi et. al., \emph{Phys. Rev.} {\bf D57}, 
589, 1998. 

\bibitem{slept} See M. Drees and M. M. Nojiri in Ref.[12]. 

\bibitem{parida} A. Datta, M. Guchait and N. Parua, \emph{Phys. Lett.}  
{\bf B395}, 54 (1997); A. Datta, Aseshkrishna Datta and M. K. Parida,
\emph{Phys. Lett.} {\bf B431}, 347 (1998). 
\bibitem{aleph} ALEPH Collaboration, R. Barate et al., \emph{Phys. Lett.} {\bf B407}, 377,
1997.

\bibitem{sally} G. L. Kane and J. P. Leveille, \emph{Phys. Lett.} {\bf B112},
227 (1982); P. R. Harrison and C. H. Llewellyn-Smith, \emph{Nucl. Phys.} {\bf  B213},
223 (1983) [Err: \emph{Nucl. Phys.} {\bf B223}, 542 (1983)]; S. Dawson and 
E. Eichten and C. Quigg, \emph{Phys. Rev.} {\bf D31}, 1581 (1985); E. Reya
and D. P. Roy, \emph{Phys. Rev.} {\bf D32}, 645 (1985). 

\bibitem{spira} W. Beenakker, M. Krammer, T. Plehn, M. Spira and P. M. Zerwas, 
\emph{Nucl. Phys.} {\bf B515}, 3 (1998).

\bibitem{pdflib} H. Plothow-Besch, PDFLIB, \emph{Comp. Phys. Comm.} {\bf 75}
396(1993). 

\bibitem{ttop} CDF Collaboration, F. Abe et al., \emph{Phys. Rev. Lett.} 
{\bf 79}, 1992(1997); F. Abe et al., \emph{Phys. Rev. Lett.} {\bf 80}, 2767 
(1998).

\end{thebibliography}
\end{document}